\documentclass[12pt]{article}
\pdfoutput=1
\usepackage{graphicx,amsfonts,amsthm,amsmath,amssymb,hyperref,color,yfonts,cite,array,caption}
\newcommand{\Tr}{{\rm Tr}}
\newcommand{\bea}{\begin{eqnarray}\displaystyle}
\newcommand{\eea}{\end{eqnarray}}

\begin{document}
\makeatletter
\@addtoreset{equation}{section}
\makeatother
\renewcommand{\theequation}{\thesection.\arabic{equation}}
\vspace{1.8truecm}

{\LARGE{ \centerline{\bf Massive fields in AdS from Constructive Holography}}}

\vskip.5cm 

\thispagestyle{empty} 
\centerline{{\large\bf Robert de Mello Koch$^{a,b,d}$\footnote{{\tt robert@zjhu.edu.cn}}, Pratik Roy$^{b,d}$\footnote{{\tt roy.pratik92@gmail.com}} and
 Hendrik J.R. Van Zyl a$^{c,d}$\footnote{\tt hjrvanzyl@gmail.com}}}
\vspace{.8cm}
\centerline{{\it $^{a}$School of Science, Huzhou University, Huzhou 313000, China,}}
\vspace{.8cm}
\centerline{{\it $^{b}$School of Physics and Mandelstam Institute for Theoretical Physics,}}
\centerline{{\it University of the Witwatersrand, Wits, 2050, }}
\centerline{{\it South Africa }}
\vspace{.8cm}
\centerline{{\it $^{c}$The Laboratory for Quantum Gravity \& Strings,}}
\centerline{{\it Department of Mathematics \& Applied Mathematics,}}
\centerline{{\it University of Cape Town, Cape Town, South Africa}}
\vspace{.8cm}
\centerline{{\it $^{d}$ The National Institute for Theoretical and Computational Sciences,}} \centerline{{\it Private Bag X1, Matieland, South Africa}}

\vspace{1truecm}

\thispagestyle{empty}

\centerline{\bf ABSTRACT}

\vskip.2cm 
Collective field theory offers a constructive framework for exploring the AdS/CFT duality. In this article, we focus on constructing rotations within the light-front quantized collective field theory for the full set of spatial coordinates in the dual bulk AdS spacetime. Two intricate aspects require attention: how rotations involving the emergent holographic coordinate are implemented, and how rotations that involve the spatial coordinates participating in the construction of the light-cone coordinates $X^{\pm}$ are realized. Our construction is in agreement with Metsaev's construction directly in the gravity theory. Additionally, we derive the eigenfunctions of the AdS mass operator, which dictate the GKPW rule for the emergent higher-dimensional theory.

\setcounter{page}{0}
\setcounter{tocdepth}{2}
\newpage
\tableofcontents
\setcounter{footnote}{0}
\linespread{1.1}
\parskip 4pt

{}~
{}~

\section{Introduction}

Collective field theory \cite{Jevicki:1979mb,Jevicki:1980zg} offers a systematic approach to the large $N$ expansion of gauge field theories by expressing the dynamics of the theory in terms of gauge invariant collective fields. In this framework, the loop expansion of the field theory corresponds to the $1/N$ expansion. Given that the dynamics of many non-Abelian gauge theories, including QCD, are strongly coupled at low energies, developing efficient methods for performing the large $N$ expansion \cite{tHooft:1973alw} is a significant and ongoing challenge. Collective field theory is highly relevant for addressing this challenge, and it is one of the primary motivations for the development of the method \cite{Jevicki:1980zq}.

Collective field theory has also proven to be a powerful approach in exploring gauge theory/gravity dualities. A clear paradigmatic example is the application of collective field theory to the dynamics of a single hermitian matrix, leading to a string field theory for the $c=1$ string \cite{Das:1990kaa}. This string field theory has successfully passed many highly non-trivial tests, demonstrating that the collective field theory of matrix quantum mechanics does reproduce the gravitational dynamics of the dual string theory.

In the context of the AdS/CFT correspondence \cite{Maldacena:1997re,Gubser:1998bc,Witten:1998qj}, the application of collective field theory, as envisioned by Das and Jevicki \cite{Das:2003vw}, has become a prototypical example of constructive approaches to holography. When applying light-front quantization to the free O($N$) vector model \cite{Klebanov:2002ja,Sezgin:2002rt} using bilocal collective fields there emerges a detailed and precise match \cite{deMelloKoch:2010wdf,deMelloKoch:2023ngh} to higher spin gravity \cite{Vasiliev:1990en,Vasiliev:2003ev}. The equal $x^+$ bilocals systematically reduce the dynamics of the vector model conformal field theory by eliminating the $+$ polarizations of currents. This reduction mirrors the light cone gauge choice in the dual higher spin gravity, which sets all components of the spinning gauge field with $+$ polarizations to zero. Consequently, a one-to-one correspondence between the degrees of freedom in higher spin gravity and those in the singlet sector of the conformal field theory is established \cite{deMelloKoch:2023ngh}. This approach allows for the precise identification of gravitational degrees of freedom and their dynamics starting from the conformal field theory, with no prior gravitational input. Thus, the method is genuinely constructive.

The mapping involves both a formula that relates the coordinates of the conformal field theory to those of the bulk spacetime in the gravity dual, and a formula that relates the fields in the two descriptions. This map is determined entirely by conformal symmetry, as we now explain. In the collective field theory description, the generators of the conformal group are given by the coproduct of the usual generators defined in the free conformal field theory. The construction of the generators in the dual gravity description is more involved but has been carried out in detail in an impressive series of papers \cite{Metsaev:1999ui,Metsaev:2003cu,Metsaev:2005ws,Metsaev:2013kaa,Metsaev:2015rda,Metsaev:2022ndg}. In the gravitational context, we fix to the light cone gauge and solve the associated constraints. A complication arises because conformal transformations do not preserve the light cone gauge choice. Thus the conformal generators are evaluated using the Killing vectors associated with the transformations and composing with compensating gauge transformations that restore the gauge after the conformal transformation \cite{Metsaev:1999ui}. Requiring that the resulting generators on the two sides agree determines the map relating coordinates and the identification of fields.

The resulting theory reproduces the correct equations of motion \cite{deMelloKoch:2014vnt}, provides a construction of bulk fields that is in manifest agreement with entanglement wedge reconstruction \cite{deMelloKoch:2021cni} and offers a concrete realization \cite{deMelloKoch:2022sul} of the holography of information \cite{Laddha:2020kvp}. Demonstrating the correctness of the equations of motion includes showing that the boundary values of the field reproduce the correct GKPW rule after transforming to the light cone gauge \cite{Mintun:2014gua}.

The collective field theory description is highly redundant because the collective fields are over complete. It has been argued in \cite{deMelloKoch:2022sul} that this redundancy is the key feature responsible for the holographic nature of the collective field theory description. Thus, this redundancy is a essential feature allowing the collective field theory to reproduce gravitational dynamics.

A number of interesting extensions of this work have been considered. For an equal time approach to the vector model/ higher spin duality see \cite{deMelloKoch:2014vnt}. The mapping for a conformal field theory with an arbitrary dimension $d$ was achieved in \cite{Jevicki:2011ss,Mintun:2014gua,deMelloKoch:2024juz}. Approaches based on a two time bilocal description have appeared in \cite{deMelloKoch:2018ivk,Aharony:2020omh}. The finite temperature conformal field theory has also been studied in \cite{Jevicki:2015sla,Jevicki:2017zay,Jevicki:2021ddf}. This work is particularly interesting given that the dual spacetimes have horizons. For studies of the collective feld theory defined by the interacting IR fixed point see \cite{Mulokwe:2018czu,Johnson:2022cbe}.

Another interesting direction, most relevant for this paper, is the extension to free matrix models \cite{deMelloKoch:2024ewt,deMelloKoch:2024otg,deMelloKoch:2024nfj}. The description of the singlet sector of a free hermitian matrix model in $d$ dimensions is achieved using the $k$-local collective fields given by
\bea
\sigma_k(t,\vec{x}_1,\vec{x}_2,\cdots,\vec{x}_k)&=&\Tr (\phi(t,\vec{x}_1)\phi(t,\vec{x}_2)\cdots\phi(t,\vec{x}_k))\label{kloccollfld}
\eea
where the field $\phi^a_b(t,\vec{x})$ $a,b=1,2,\cdots N$ is an $N\times N$ hermitian matrix. In addition we have the conjugate momentum
\bea
\Pi_k(t,\vec{x}_1,\vec{x}_2,\cdots,\vec{x}_k)&=&-{1\over i}{\delta\over\delta \sigma_k(t,\vec{x}_1,\vec{x}_2,\cdots,\vec{x}_k)}
\eea
The collective Hamiltonian is given by rewriting the theory in terms of invariant fields. For the conjugate momentum, the transition from the original momentum $\pi$ to the momentum conjugate to the collective field (\ref{kloccollfld}) is accomplished through the use of the functional chain rule\footnote{The explicit appearance of gauge indices $a$ and $b$ on the right hand side is a consequence of the fact that we are rewriting $\pi^b_a(t,\vec{x})$ which itself is not gauge invariant. Any gauge invariant expression, including the kinetic term, $\Tr(\pi(t,\vec{x})\pi(t,\vec{x}))$ is expressed entirely in terms of the collective fields $\sigma_k(t,\vec{x}_1,\vec{x}_2,\cdots,\vec{x}_k)$ and $\Pi_k(t,\vec{x}_1,\vec{x}_2,\cdots,\vec{x}_k)$ with no gauge indices appearing in the final expression.}
\bea
\pi^b_a(t,\vec{x})&=&-{1\over i}{\delta\over\delta\phi^a_b(t,\vec{x})}\cr\cr
&=&-{1\over i}\sum_{k}\int d^{d-1}y_1\cdots \int d^{d-1}y_k {\delta \sigma_k(t,\vec{y}_1,\vec{y}_2,\cdots,\vec{y}_k)\over\delta \phi^a_b(t,\vec{x})}{\delta\over\delta \sigma_k(t,\vec{y}_1,\vec{y}_2,\cdots,\vec{y}_k)}\cr\cr
&=&\sum_{k}\int d^{d-1}y_1\cdots \int d^{d-1}y_k {\delta \sigma_k(t,\vec{y}_1,\vec{y}_2,\cdots,\vec{y}_k)\over\delta \phi^a_b(t,\vec{x})}\Pi_k(t,\vec{y}_1,\vec{y}_2,\cdots,\vec{y}_k)
\eea
This change of variables is accompanied by a non-trivial Jacobian which produces an infinite sequence of interaction vertices. For details see the original papers \cite{Jevicki:1979mb,Jevicki:1980zg}. The usual equal time commutator\footnote{The terms $\cdots$ in this expression are related by a cyclic permutation $\sigma\in\mathbb{Z}_k$. These extra terms are needed to account for the cyclic symmetry of the collective fields.}
\bea
[\Pi_k(t,\vec{y}_1,\vec{y}_2,\cdots,\vec{y}_k),\sigma_{k'}(t,\vec{x}_1,\vec{x}_2,\cdots,\vec{x}_{k'})]&=&-i\delta_{kk'}\prod_{i=1}^k\delta (\vec{x}_i-\vec{y}_i)+\cdots
\eea
makes it clear that we have quantized an independent degree of freedom living at each position in the $k(d-1)$ dimensional space with coordinates $\vec{x}_1,\cdots,\vec{x}_k$. If the collective field theory is to define a physically coherent framework, this complete set of $k(d-1)$ spatial coordinates of the $k$-local collective field must have a sensible interpretation as coordinates of a space in the dual gravity description. Demonstrating that this is indeed the case has been achieved in \cite{deMelloKoch:2024ewt,deMelloKoch:2024otg,deMelloKoch:2024nfj} with the result that the collective field is defined on a spacetime that is a product of AdS$_{d+1}$ times a product of spheres $S^{k-1}\times S^{(d-2)(k-2)}\times S^{d-3}$. The coefficients of a harmonic expansion of the collective fields on the $S^{(k-1)}\times S^{(d-2)(k-2)}\times S^{d-3}$ factor are the bulk fields of the dual gravity theory so that we are describing a structure responsible for organizing the spinning fields in the gravity theory, and the primary fields in the conformal field theory. For $k=2$ the collective field gives rise to a single scalar primary, and a tower of conserved spinning currents. The spin runs over the even integers and there is only one primary of a given spin. Since the currents are conserved they all belong to short representations of the conformal group. In the dual gravity we obtain a single scalar and a tower of massless spinning gauge fields. The $k=2$ sector is in common with the O($N$) vector model. In the matrix model we also have collective fields with $k>2$. None of the primary operators packaged by these collective fields are conserved and they all belong to usual (i.e. not short) conformal representations. In the dual gravity these are massive fields. Taken together, the collective field theory of these fields gives a description of the tensionless limit of a string theory\cite{Mikhailov:2002bp}. We will work in the $N=\infty$ limit so that our gravity fields are not interacting.

For the remainder of this paper, we employ light-front quantization. This choice is motivated by the fact that, in this framework, the requirement of bulk locality effectively determines much of the holographic mapping \cite{deMelloKoch:2024juz}. The relevant results are reviewed in Section \ref{Bcoords}. To frame the questions we consider, recall that in bulk AdS$_{d+1}$ spacetime, light-front quantization uses the time $T$ and a spatial coordinate $X^{d-1}$ to define the light cone coordinates $X^\pm={1\over\sqrt{2}}(X^{d-1}\pm T)$. Apart from the extra holographic coordinate $Z$, there are $d-2$ spatial coordinates $X^a$ ($a=1,2,...,d-2$) transverse to the light cone. In this light-front description, rotations mixing the $X^a$ are manifest symmetries described by the group SO($d-2$), which transparently matches up with an equivalent subgroup in the conformal field theory. Understanding the SO($d-1$) group, within the collective field theory, that rotates  $X^a$ and $Z$ is more challenging. It is even more non-trivial to obtain the full SO($d$) group that mixes the complete set of spatial coordinates $X^a$, $X^{d-1}$ and $Z$. Explaining how this SO($d$) group emerges in the collective field theory is the main result of this paper. With this understanding, we also derive the eigenfunctions of the AdS mass operator, which determine the boundary behaviour of the constructed bulk fields as $Z\to0$. In other words, we determine the complete GKPW rule for the constructed bulk fields.

Our construction of the emergent SO($d$) symmetry in collective field theory, as with our previous constructions of holography for the free matrix \cite{deMelloKoch:2024ewt,deMelloKoch:2024otg,deMelloKoch:2024nfj}, relies heavily on the remarkable results of Metsaev \cite{Metsaev:1999ui}. In a completely gauge fixed light cone gauge description of spinning fields in AdS, the realization of the SO($d-1$) rotations that mix $X^a$ with $Z$ is connected to a description of the so($d-1$) algebra in stereographic coordinates \cite{Metsaev:1999ui}. Our first result in Section \ref{Interp} demonstrates that collective field theory also realizes this algebra using stereographic coordinates. A curious twist on the analysis of \cite{Metsaev:1999ui} is that we find $k$ copies of the algebra participate, but there are constraints relating the different copies. To understand how the full so($d$) algebra is realized, \cite{Metsaev:1999ui} introduces an so($d-1$,1) algebra. A very similar construction is possible in the collective field theory, with all $k$ copies of the so($d-1$) algebra participating. This again confirms that collective field theory reconstructs the dual gravitational dynamics in complete detail. With these new insights, and drawing once more on \cite{Metsaev:1999ui}, we present the eigenfunctions of the AdS mass operator in Section \ref{BC}. In Section \ref{conclusions}, we draw some conclusions and highlight several intriguing directions for future research. In the Appendices we demonstrate that the eigenfunctions of the AdS mass operator that we have constructed are in agreement with existing results for special cases available in the literature.

\section{Bulk coordinates}\label{Bcoords}

In this section we review the key features of the holographic mapping obtained in \cite{deMelloKoch:2024nfj}, which builds on the earlier results \cite{deMelloKoch:2024juz,deMelloKoch:2024ewt,deMelloKoch:2024otg}. The $k$-local collective field 
\bea
\sigma_k(x^+,x_1^-,x^a_1,x_2^-,x^a_2,\cdots,x_k^-,x^a_k)&=&\Tr (\phi(x^+,x_1^-,x^a_1)\phi(x^+,x_2^-,x^a_2)\cdots\phi(x^+,x_k^-,x^a_k))\nonumber
\eea
develops a large $N$ expectation value, denoted by $\sigma_k^0(x^+,x_1^-,x^a_1,\cdots,x_k^-,x^a_k)$. We expand about this large $N$ value to define a fluctuation as follows
\bea
\sigma_k(x^+,x_1^-,x^a_1,\cdots,x_k^-,x^a_k)&=&\sigma_k^0(x^+,x_1^-,x^a_1,\cdots,x_k^-,x^a_k)+{1\over N^{k\over 2}}\eta_k(x^+,x_1^-,x^a_1,\cdots,x_k^-,x^a_k)\cr
&&\label{klocfluct}
\eea
The coefficient of the fluctuation in the above equation ensures that the two point function of $\eta_k$ is order 1 as $N\to\infty$\footnote{It is useful to recall that for odd $k$ $\sigma_k^0(x^+,x_1^-,\vec{x}_1,\cdots,x_k^-,\vec{x}_k)$ vanishes while for even $k$ it is of order $N^{{k\over 2}+1}$. These large $N$ expectation values are easily evaluated using free field theory.}. It is the fluctuation $\eta_k$ that maps to the bulk gravity field $\Phi$. The derivation of the holographic map that identifies the fields as well as the coordinates of the collective field theory with those of the dual gravitational theory, is most easily carried out \cite{deMelloKoch:2021cni} after performing a Fourier transform. In the bulk\footnote{We will consistently use capital letters for the coordinates of the AdS bulk spacetime and little letters for the coordinates of the collective field theory.} this Fourier transform replaces $X^-$ with $P^+$, while in the boundary the corresponding Fourier transform replaces $x^-$ with $p^+$. The relation between the fields of the collective field theory and those of the bulk gravity takes the form
\bea
\Phi(X^+,P^+,X^a,Z,\{\alpha_i\})
=\mu(x^a_i,p_i^+)\eta_k(x^+,p_1^+,x^a_1,\cdots,p_k^+,x^a_k)\label{relflds}
\eea
with \cite{deMelloKoch:2024nfj}
\bea
\mu(x^a_i,p_i^+)&=&Z^{3-d\over 2}\,(P^+ Z)^{{k\over 2}(d-2)-1}\,\Big(\prod_{i=1}^k p_i^+\Big)^{4-d\over 2}\label{FormForMu}
\eea
In the above formula $\{\alpha_i\}$ are simply a place holder for the extra coordinates that arise from the $k$-local collective field. We do not spell these out in (\ref{relflds}) as a number of different choices will be useful at different points in the analysis. We start in Section \ref{Interp} with geometrically motivated coordinates $\{\alpha_i\}=\{\beta^a_q,(\hat{n}_{P^+})_q\}$. We then change to coordinates $\{\alpha_i\}=\{\zeta^Z_q,\zeta^a_q\}$ which transform as a vector under the SO($d-1$) subgroup rotating the $X^a,Z$ coordinates. Finally we consider coordinates $\{\alpha_i\}=\{\xi^a_q,\rho_q\}$ which are most useful for describing the complete set of SO($d$) rotations. These different sets of coordinates are all introduced and carefully defined in Section \ref{Interp}.

The formula expressing the bulk coordinates in terms of the coordinates of the collective $d$ dimensional $k$-local collective field are determined by the requirement of bulk locality as explained in \cite{deMelloKoch:2024juz}. The formulas for the bulk coordinates are
\bea
X^+&=&x^+\qquad\qquad P^+\,\,=\,\,\sum_{i=1}^k p_i^+\qquad\qquad Z\,\,=\,\,{\sqrt{\sum_{a=1}^{d-2}\sum_{i=1}^k p_i^+ (v_i^a)^2}\over (\sum_{l=1}^k p_l^+)^{3\over 2}}\cr\cr\cr
X^a&=&{\sum_{i=1}^k p_i^+ x_i^a\over \sum_{l=1}^k p_l^+}\qquad a=1,2,\cdots,d-2 \label{BulkCoords}
\eea
where 
\bea
v_i^a&=&\sum_{l=1}^k p_l^+ (x_i^a-x_l^a) \label{defvai}
\eea
We use $Z$ to denote the extra bulk holographic coordinate. The $d+1$ coordinates $X^+,P^+,X^a,Z$ parametrize the AdS$_{d+1}$ spacetime. A noteworthy feature of the formula for $Z$ is that it implies that the boundary $Z=0$ is only reached when all fields in the $k$-local become coincident
\bea
x_i^a&=&x_j^a\qquad\qquad a=1,2,\cdots,d-2\qquad i,j=1,2,\cdots,k
\eea
As discussed in \cite{deMelloKoch:2023ylr} this is a signal of the holographic nature of the collective field theory. Bulk locality also fixes the form of the spin generators \cite{deMelloKoch:2024juz}. The result is
\bea
M^{aZ}&=&{1\over Z(P^+)^2}\sum_{b=1}^{d-2}\sum_{j=1}^k v^b_j\left(v^a_j{\partial\over\partial x^b_j}-v^b_j{\partial\over\partial x^a_j}\right)\cr\cr
&+&{1\over 2Z(P^+)^2}\sum_{j,l=1}^k\left({p_l^+\over P^+}\sum_{b=1}^{d-2}(v^b_jv^b_j+v^b_lv^b_l){\partial\over\partial x^a_j}-2p_j^+p_l^+v^a_j{\partial\over\partial p^+_j}\right)\label{tzlcspins}
\eea
\bea
M^{ab}={1\over P^+}\sum_{l=1}^k\left(v_l^a{\partial\over\partial x_l^b}-v_l^b{\partial\over\partial x_l^a}\right)\label{tlcspins}
\eea
The final spin generators $M^{-a}$ are determined using the relation \cite{Metsaev:1999ui}
\bea
M^{-a}&=&-M^{a-}\,\,=\,\,M^{aZ}{\partial_Z\over P^+}+\sum_{b=1}^{d-2}M^{ab}{\partial_b\over P^+}-{1\over 2ZP^+}\sum_{b=1}^{d-2}[M^{Zb},M^{ba}]
\eea
The bulk gravity equation of motion takes the form\cite{Metsaev:2003cu}
\bea
\left(2P^+P^-+\sum_{a=1}^{d-2}\partial_{X^a}\partial_{X^a}+\partial_Z^2-{A\over Z^2}\right)\Phi=0
\eea
where $A$ is the AdS mass operator. In terms of the collective field theory coordinates the AdS mass operator takes the form
\bea
A&=&-{1\over 4(k-3)!}\sum_{i_1i_2\cdots i_{k-3}=1}^k\sum_{a=1}^{d-2}\kappa_{aa;i_1i_2\cdots i_{k-3}}\sum_{b=1}^{d-2}\kappa_{bb;i_1i_2\cdots i_{k-3}}\cr\cr
&&-{1\over 2(k-3)!}\sum_{a,b=1}^{d-2}\sum_{i_1i_2\cdots i_{k-3}=1}^k \left(\kappa_{ab;i_1i_2\cdots i_{k-3}}\kappa_{ab;i_1i_2\cdots i_{k-3}}-\kappa_{aa;i_1i_2\cdots i_{k-3}}\kappa_{bb;i_1i_2\cdots i_{k-3}}\right)\cr\cr\cr
&&-{1\over 2}\sum_{a,b=1}^{d-2}M_{ab}^2+{((k-1)d-2k+1)((k-1)d-2k-1)\over 4}
\label{AdSMassOp}
\eea
where
\bea
\kappa_{aa;i_1i_2\cdots i_{k-3}}&=&Z\sum_{i_{k-2},i_{k-1},i_k=1}^k\sum_{a=1}^{d-2}\epsilon_{i_1 i_2\cdots i_k}{\beta^a_{i_{k-2}}\sqrt{p^+_{i_{k-1}}}-\beta^a_{i_{k-1}}\sqrt{p^+_{i_{k-2}}}\over\sqrt{p^+_{i_k}}}{\partial\over\partial x^a_{i_k}}
\eea
\bea
\beta^a_i&=&{\sqrt{p_i^+}v^a_i\over Z (\sum_{l=1}^k p_l^+)^{3\over 2}}\label{defbeta}
\eea
and
\bea
\kappa_{ab;i_1i_2\cdots i_{k-3}}&=&{1\over 2}\left[ M_{ab},\kappa_{aa;i_1i_2\cdots i_{k-3}}\right]
\eea

\section{Emergent SO($d$)}\label{Interp}

In this section we construct the group of SO($d$) rotations that mix the complete set of spatial coordinates of the emergent AdS$_{d+1}$ spacetime. The generators of the group take the form
\bea
{\cal L}^{AB}&=&L^{AB}+M^{AB}
\eea
where $L^{AB}$ is the orbital angular momentum and $M^{AB}$ is the spin angular momentum. Formulas for the $L^{AB}$ are easily given in terms of the bulk coordinates. We are interested in the realization of the above generators when they act on general massive spinning fields in the bulk AdS$_{d+1}$ spacetime that emerges from a collective field theory treatment of the free matrix conformal field theory. In this case the construction of the spin generators $M^{AB}$ is non-trivial. This is the problem that is solved in this section. These spin generators have also been constructed directly in the gravity theory using a complete light cone gauge fixing of the theory in \cite{Metsaev:1999ui}. The gravity results of \cite{Metsaev:1999ui} are in complete harmony with the collective field theory results obtained in this section.

The coordinates the collective field are a light cone momentum and a vector collecting coordinates transverse to the light cone
\bea
(\vec{x}^a)_i\,\,\equiv\,\, x^a_i\quad\qquad\qquad(\vec{p}^+)_i\,\,=\,\,p^+_i\qquad\qquad\quad i&=&1,\cdots,k\cr
 a&=&1,\cdots,d-2
\eea
The vectors $\vec{x}^a$ and $\vec{p}^+$ are elements of the $k$-dimensional vector space $V_k$. We can think of the $k$-local collective field as describing a composite system with $k$-constituents. The bulk light cone momentum is a sum of the constituent light cone momenta and the centre of mass position is given by the dot product of the light cone momentum vector $\vec{p}^+$ and the constituent position vectors as follows\footnote{Since $X^a$ is identical to the formula for the centre of mass of a collection of $k$ particles, with the $i$th particle having a mass $p^+_i$, we often refer to $X^a$ as the centre of mass coordinate.}
\bea
X^a&=&{\vec{x}^a\cdot \vec{p}^+\over P^+}\qquad\qquad P^+\,\,\equiv\,\,\sum_{i=1}^k p^+_i
\eea
This observation motivates a geometrical description: motions of the constituent positions $x^a_i$ that are orthogonal to $\vec{p}^+$ correspond to internal motions with the bulk position $X^a$ inert. Indeed, we can rewrite the positions $x^a_i$ as follows
\bea
x^a_i&=&X^a+{v_i^a\over P^+}\qquad\qquad\qquad v_i^a\,\,=\,\,\sum_{l=1}^k p_l^+(x_i^a-x_l^a)
\eea
where the vector $\vec{v}^a$ has two important properties:
\begin{itemize}
\item[1.] It is orthogonal to $\vec{p}^+$
\bea
\vec{v}^a\cdot\vec{p}^+&=&0\label{orthogconst}
\eea

\item[2.] Under a simultaneous translation $x^a_i\to x^a_i+c$ the centre of mass shifts as $X^a\to X^a+c$ but $\vec{v}^a$ is invariant.
\end{itemize} 

Consequently changing the $\vec{v}^a$ produces an ``internal motion'' in the sense that the position $\vec{x}^a$ are changed in such a way that the bulk position $X^a$ (i.e. the centre of mass) is invariant. To develop this geometrical picture further introduce the unit vector
\bea
(\hat{n}_{\vec{p}^+})_i&\equiv&{\sqrt{p_i^+}\over\sqrt{P^+}}
\eea
and a rescaled version of $v_i^a$
\bea
\beta_i^a&=&{\sqrt{p_i^+}v_i^a\over Z(P^+)^{3\over 2}}\label{defbeta}
\eea
The factor of $\sqrt{p_i^+}$ ensures that orthogonality of $\vec{v}^a$ and $\vec{p}^+$ translates into orthogonality of $\vec{\beta}^a$ and $\hat{n}_{\vec{p}^+}$ i.e.
\bea
\vec{p}^+\cdot\vec{v}^a\,\,=\,\,0\qquad&\Rightarrow&\qquad 
\hat{n}_{\vec{p}^+}\cdot\vec{\beta}^a\,\,=\,\,0
\eea
The extra factors involving $P^+$ and $Z$ appearing in the definition of $\beta^a_i$ ensure that
\bea
\sum_{a=1}^{d-2}\vec{\beta}^a\cdot\vec{\beta}^a&=&\sum_{a=1}^d\sum_{i=1}^k\beta^a_i\beta^a_i\,\,=\,\,1\label{betaconstr}
\eea
Taken together these observations give a natural geometrical motivation for the definition of $\beta^a_i$ given in (\ref{defbeta}). Notice that we can write the positions $x^a_i$ in terms of $\beta^a_i$ as follows
\bea
x^a_i&=&X^a+\beta_i^a{\sqrt{P^+}Z\over\sqrt{p_i^+}}
\eea
The quantities $\beta^a_i$ and the unit vector $\hat{n}_{\vec{p}^+}$, as well as the bulk coordinates $X^+,P^+,X^a$ and $X^{d-1}$ provide a complete set of coordinates for the $k$-local collective field.

A natural set of operators that we can define at this point are given by 
\bea
M^{ab}_{jk}&=&\beta^a_j{\partial\over\partial\beta^b_k}-\beta^b_k{\partial\over\partial\beta^a_j}\label{MabGens}
\eea
These operators generate ``rotations'' of the $\beta^a_i$ that in general mix both the spacetime index $a$ as well as the index $i$ which labels the different constituent fields appearing in the $k$-local collective field. In terms of these operators the AdS mass operators takes a strikingly simple form
\bea
A&=&-{1\over 2}\sum_{i,j=1}^k\sum_{a,b=1}^{d-2}M^{ab}_{ij}M^{ab}_{ij}+{((k-1)(d-2)-2)^2-1\over 4}\label{simplsmassop}
\eea
A comment is in order. As a consequence of the fact that $\vec{\beta}^a\cdot\hat{n}_{P^+}=0$ we have\footnote{The $a$ and $i$ indices are raised and lowered with the Kronecker delta.}
\bea
{\partial\over\partial\beta^a_j}\,\beta^b_i\,\,=\,\,\widehat{\delta}^j_i\delta^b_a &=&(\delta^{j}_i-(\hat{n}_{P^+})^j(\hat{n}_{P^+})_i)\delta^b_a
\eea
where $\widehat{\delta}^j_i$ acts as the identity in the subspace of $V_k$ that is orthogonal to $\hat{n}_{P^+}$. It is also easy to check\footnote{For example, $\beta^a_i$ is in the spin $s=1$ representation. The value of the quadratic Casimir ${1\over 2}\sum_{a,b}\sum_{ij}M^{ab}_{ij}M^{ab}_{ij}$ acting on the spin $s$ representation (totally symmetric and traceless tensors with $s$ indices) is $s(s+K-2)$ for SO($K$). We have evaluated the action of the quadratic Casimir on $\beta^a_i$ to fix $K$.} that the group generated by the (\ref{MabGens}) is given by SO($(k-1)(d-2)$).

Using the definition of $\beta^a_i$ given in (\ref{defbeta}) as well as the spin generators given in (\ref{tlcspins}) we easily obtain
\bea
M^{ab}\beta^c_i&=&\delta^{bc}\beta^a_i-\delta^{ac}\beta^b_i
\eea
These relations show that $\beta^a_i$ transforms as an SO($d-2$) vector with index $a$. Motivated by this observation we can demonstrate that
\bea
M^{ab}&=&\sum_{j=1}^kM^{ab}_{jj}\,\,=\,\,\sum_{j=1}^k\Big(\beta^a_j{\partial\over\partial\beta^b_j}-\beta^b_j{\partial\over\partial\beta^a_j}\Big)
\eea
Next, using the result (\ref{tzlcspins}) we verify that
\bea
M^{aZ}(\hat{n}_{P^+})_i&=&-{1\over 2}\beta^a_i\cr\cr
M^{aZ}\beta^c_q&=&{\beta^a_q\beta^c_q\over 2(\hat{n}_{P^+})_q}+{\delta^{ac}\over 2}\Big((\hat{n}_{P^+})_q-{1\over(\hat{n}_{P^+})_q}\sum_{b=1}^{d-2}\beta^b_q\beta^b_q\Big)\label{MaZact}
\eea
With a little more work we can prove that
\bea
M^{aZ}&=&{1\over 2}\sum_{i=1}^k\left((\hat{n}_{P^+})_i{\partial\over\partial\beta^a_i}-\beta^a_i{\partial\over\partial (\hat{n}_{P^+})_i}\right)\cr\cr
&&+{1\over 2}\sum_{i=1}^k\sum_{b=1}^{d-2}{\beta^b_i\over (\hat{n}_{P^+})_i}\left(\beta^a_i{\partial\over\partial\beta^b_i}-\beta^b_i{\partial\over\partial\beta^a_i}\right)
\eea
The first term is a rotation between $(\hat{n}_{P^+})_i$ and $\beta^a_i$, while the second is a $\beta$ dependent rotation mixing $\beta^a_i$ and $\beta^b_i$.

Some experience with the relations (\ref{MaZact}) suggests that we introduce the quantities
\bea
\zeta^Z_q&=&(\hat{n}_{P^+})_q^2-\sum_{b=1}^{d-2}\beta^b_q\beta^b_q\,\,=\,\,{p_q^+\over P^+}-\sum_{b=1}^{d-2}{p_q^+v_q^bv_q^b\over Z^2 (P^+)^3}\cr\cr
\zeta^a_q&=&-2(\hat{n}_{P^+})_q\beta^a_q\,\,=\,\,-2{p_q^+v_q^a\over Z(P^+)^2}\label{NiceStTrans}
\eea
which obey
\bea
M^{aZ}\zeta_q^Z&=&\zeta_q^a\qquad\qquad M^{aZ}\zeta_q^b\,\,=\,\,-\delta^{ab}\zeta_q^Z\cr\cr
M^{ab}\zeta_q^Z&=&0\qquad\qquad M^{ab}\zeta^c_q\,\,=\,\,\delta^{bc}\zeta^a_q-\delta^{ac}\zeta^b_q
\eea
This demonstrates that the $k$ $d-1$ dimensional vectors with components given by $\zeta_q^I\equiv\{\zeta^a_q,\zeta^Z_q\}$, $q=1,2,\cdots,k$ transform as vectors of the SO($d-1$) group that mixes the $X^a$ coordinates and the extra holographic coordinate $Z$. This demonstrates that rotations mixing the extra holographic coordinate do indeed emerge naturally from the collective field theory.

Notice that the $\{\zeta^a_q,\zeta^Z_q\}$ are an invertible function of the $\beta^a_q$s and the $(\hat{n}_{P^+})_q$s, so that, together with the bulk coordinates they provide a complete set of coordinates for the $k$-local collective field. These quantities are not all independent as they obey some interesting constraints. Indeed we find that
\bea
\sum_{q=1}^k\zeta_q^a&=&0
\eea
as a consequence of (\ref{orthogconst}) and
\bea
\sum_{q=1}^k\zeta_q^Z&=&0
\eea
as a consequence of the formula for $Z$ and the definition of $P^+$, given in (\ref{BulkCoords}). Finally, the constraint (\ref{betaconstr}) implies that
\bea
\sum_{q=1}^k\sum_{a=1}^{d-2}{\zeta^a_q\zeta^a_q\over 4 (n_{P^+})^2_q}&=&1
\eea
Altogether this is a total of $d$ constraints.

Recall that $a$ runs over $d-2$ values. There is a total of $(d-1)k$ variables $\zeta_q^I$. After solving the constraints, we see that $(d-1)k-d$ of them are independent. The $k$-local collective field (\ref{klocfluct}) contains a total of $1+k(d-1)$ coordinates. We need $d+1$ coordinates to describe the dual AdS$_{d+1}$ spacetime, so there are a total of $k(d-1)-d$ ``internal'' coordinates. These are provided by the $\zeta^I_q$ given above. The advantage of this new description as compared to the description using $p^+_i$ and $\beta^a_i$, is that we now have a correct understanding of how these parameters transform under the spacetime SO($d-1$) transformations.

To obtain some insight into the SO($d-1$) group we have uncovered above, introduce $\xi^a_i=-\beta^a_i/(\hat{n}_{P^+})_i$ so that
\bea
\zeta^Z_q&=&(\hat{n}_{P^+})_q^2\Big(1-\sum_{b=1}^{d-2}\xi^b_q\xi^b_q\Big)\,\,\equiv\,\,\rho_q\Big(1-\sum_{b=1}^{d-2}\xi^b_q\xi^b_q\Big)\cr\cr
\zeta^a_q&=&2(\hat{n}_{P^+})^2_q\xi^a_q\,\,=\,\, 2\rho_q\xi^a_q
\eea
This nicely matches the formula (B.3) of \cite{Metsaev:1999ui} and it demonstrates that we have obtained a description of the so($d-1$) algebra in terms of stereographic coordinates. In terms of these new coordinates we can define $k$ independent spin generators as follows
\bea
M_q^{ab}&=&\xi_q^a\partial_{\xi_q^b}-\xi_q^b\partial_{\xi_q^a}\cr\cr
M_q^{za}&=&\frac{1}{2}\Big(1-\sum_{b=1}^{d-2}\xi^b_q\xi^b_q\Big)\partial_{\xi_q^a}+\xi^a_q\Big(\sum_{b=1}^{d-2}\xi^b_q \partial_{\xi^b_q}-\rho_q\partial_{\rho_q}\Big)
\eea
These close the SO($d-1$) algebra. In terms of these coordinates, the spacetime spin generators take the following simple form
\bea
M^{ab}&=&\sum_{q=1}^k M^{ab}_q\,\,=\,\,\sum_{q=1}^k\Big(\xi_q^a\partial_{\xi_q^b}-\xi_q^b\partial_{\xi_q^a}\Big)\cr\cr
M^{za}&=&\sum_{q=1}^k M^{za}_q\,\,=\,\,\sum_{q=1}^k\Big(\frac{1}{2}\Big(1-\sum_{b=1}^{d-2}\xi^b_q\xi^b_q\Big)\partial_{\xi_q^a}+\xi^a_q\Big(\sum_{b=1}^{d-2}\xi^b_q \partial_{\xi^b_q}-\rho_q\partial_{\rho_q}\Big)\Big)
\eea

To exhibit the complete bulk SO($d$) group, which rotates $X^a$, $Z$ and $X^{d-1}$, we have to include rotations in which $X^{d-1}$ participates. Recall that this is the coordinate used to construct $X^\pm$. Metsaev has achieved this \cite{Metsaev:1999ui} by first demonstrating how to define an so($d-1$,1) algebra. Using this additional so($d-1$,1) structure it is then possible to explicitly construct the carrier space of the SO($d$) representation needed to describe a massive spinning field in the AdS bulk. A very similar construction is possible in collective field theory and this too allows us to describe the SO($d$) representations needed to describe the massive spinning fields packaged into the collective field. Our construction introduces the operators
\bea
{\sf p}_q^a&=&{\partial\over\partial\xi_q^a}\cr\cr
{\sf k}_q^a&=&-\frac{1}{2}\Big(\sum_{b=1}^{d-2}\sum_{i=1}^k \xi^b_i\xi^b_i\Big){\partial\over\partial\xi^a_q}+\xi^a_q\Big(\sum_{b=1}^{d-2}\sum_{i=1}^k\xi^b_i{\partial\over\partial\xi^b_i}-\sum_{i=1}^k \rho_i{\partial\over\partial\rho_i}\Big)\cr\cr
{\sf m}^{ab}_{ij}&=&\xi^a_i{\partial\over\partial \xi^b_j}-\xi^b_j{\partial\over\partial \xi^a_i}\cr\cr
{\sf d}&=&\sum_{b=1}^{d-2}\sum_{i=1}^k\xi^b_i{\partial\over\partial\xi^b_i}-\sum_{i=1}^k \rho_i{\partial\over\partial\rho_i}
\eea
This algebra is the Lie algebra of SO($(k-1)(d-2)+1,1$). The ${\sf m}^{ab}_{ij}$ generate the subgroup SO($(k-1)(d-2)$). These operators satisfy the following commutation relations
\bea
[{\sf d},{\sf p}^a_q]&=&-{\sf p}^a_q \qquad [{\sf d}_q,{\sf k}^a_q]\,\,=\,\,{\sf k}^a_q\cr\cr
[{\sf p}^a_q,{\sf p}^b_q]&=&0\qquad\qquad [{\sf k}^a_q,{\sf k}^b_q]\,\,=\,\,0\cr\cr
[{\sf p}^a_q,{\sf m}^{bc}_{ij}]&=&\delta^{ab}\widehat{\delta}_{iq}{\sf p}^c_j-\delta^{ac}\widehat{\delta}_{jq}{\sf p}^b_i\cr\cr
[{\sf k}^a_q,{\sf m}^{bc}_{ij}]&=&\delta^{ab}\widehat{\delta}_{iq}{\sf k}^c_j-\delta^{ac}\widehat{\delta}_{jq}{\sf k}^b_i\cr\cr
[{\sf p}^a_i,{\sf k}^b_j]&=&\delta^{ab}\widehat{\delta}_{ij}{\sf d}-{\sf m}^{ab}_{ij}\cr\cr
[{\sf m}^{ab}_{ij},{\sf m}^{cd}_{lm}]&=&\delta^{bc}\widehat{\delta}_{jl}{\sf m}^{ad}_{im}-\delta^{ac}\widehat{\delta}_{il}{\sf m}^{bd}_{jm}-\delta^{bd}\widehat{\delta}_{jm}{\sf m}^{ac}_{il}+\delta^{ad}\widehat{\delta}_{im}{\sf m}^{bc}_{jl}
\eea
Notice that, if we have an eigenstate of ${\sf d}$ with eigenvalue $\lambda$, then applying ${\sf k}^a_i$ produces another eigenstate with eigenvalue $\lambda+1$ and applying ${\sf p}^a_i$ also produces another eigenstate with eigenvalue $\lambda-1$. In this sense ${\sf k}^a_i$ and ${\sf p}^a_i$ are raising and lowering operators for ${\sf d}$. To make manifest the so($(k-1)(d-2)+1,1$) algebra, introduce the new generators
\bea
J^{ab}_{ij}&=&{\sf m}^{ab}_{ij}\cr\cr
J^{Za}_i&=&{1\over 2}{\sf p}^a_i+{\sf k}^a_i\cr\cr
J^{d-1\, a}_i&=&{1\over 2}{\sf p}^a_i-{\sf k}^a_i\cr\cr
J^{Z\, d-1}&=&{\sf d}
\eea
They close the algebra
\bea
[J^{ab}_{ij},J^{cd}_{lm}]&=&\delta^{bc}\widehat{\delta}_{jl}J^{ad}_{im}-
\delta^{ac}\widehat{\delta}_{il}J^{bd}_{jm}-\delta^{bd}\widehat{\delta}_{jm}J^{ac}_{il}+\delta^{ad}\widehat{\delta}_{im}J^{bc}_{jl}\cr\cr
[J^{aZ}_i,J^{bZ}_j]&=&-J^{ab}_{ij}\qquad\qquad
[J^{a\, d-1}_i,J^{bZ}_j]\,\,=\,\,-\delta^{ab}\widehat{\delta}_{ij}J^{d-1\, Z}\cr\cr
[J^{ab}_{ij},J^{Zc}_{l}]&=&\delta^{ac}\widehat{\delta}_{il}J^{bZ}_j-\delta^{bc}\widehat{\delta}_{jl}J^{aZ}_i\qquad\quad [J^{ab}_{ij},J^{Z \, d-1}]\,\,=\,\,0\cr\cr
[J^{ab}_{ij},J^{d-1\,c}_{l}]&=&\delta^{ac}\widehat{\delta}_{il}J^{b\, d-1}_j-\delta^{bc}\widehat{\delta}_{jl}J^{a\, d-1}_i\qquad [J^{Za}_i,J^{Z\, d-1}]\,\,=\,\,J^{d-1\, a}_i\cr\cr
[J^{d-1\, a}_i,J^{d-1\, b}_j]&=&-J^{ab}_{ij}\qquad\qquad
[J^{d-1\, a}_i,J^{Z\, d-1}]\,\,=\,\,J^{aZ}_i
\eea
The spacetime generators of SO($d-1$,1) rotations are
\bea
J^{ab}&=&\sum_{i=1}^k J^{ab}_{ii}\qquad
J^{Za}\,\,=\,\,\sum_{i=1}^kJ^{Za}_i\qquad
J^{d-1\,a}\,\,=\,\,\sum_{i=1}^k J^{d-1\,a}_i\qquad 
J^{Z\, d-1}\,\,=\,\,{\sf d}
\eea
and we have verified that they close the correct so($d-1$,1) algebra. To obtain the so($d$) algebra we Wick rotate as usual. Following Metsaev \cite{Metsaev:1999ui} a basis for the spin states needed for the description of a massive spinning field can now be constructed. For this construction we organize the carrier space of the SO($d$) irreducible representation in terms of the SO(2)$\times$SO($d-2$) subgroup of SO($d$) where SO($d-2$) is generated by $J^{ab}$ and SO(2) is generated by $J^{Z\, d-1}={\sf d}$. Clearly then, the special conformal like generator ${\sf k}^a_i={1\over 2}(J^{Za}_i-J^{d-1\,a}_i)$ will raise the SO(2) quantum numbers. This motivates us to introduce a vector $|{\bf 0}\rangle$ which is an eigenvector of the operator ${\sf d}$ with eigenvalue equal to twice the spin $s$ of the representation and it is annihilated by ${\sf p}^a_i$. This condition is imposed to obtain a finite dimensional representation. These constraints imply that $|{\bf 0}\rangle$ does not depend on the $\xi^a_i$ and that they are homogeneous in the $\rho_i$ with total degree equal to twice the spin. The space
\begin{equation}\label{kefuncs}
\Lambda\equiv\bigoplus_{\sigma=0}^{2s}\Lambda^\sigma
\qquad\qquad \Lambda^\sigma\equiv {\sf k}^{a_1}_{i_1}\ldots {\sf k}^{a_\sigma}_{i_\sigma}|{\bf 0}\rangle
\end{equation}
can be used to describe the spin degrees of freedom of a massive field. To obtain a state in $\Lambda^\sigma$ that has a definite SO($d-2$) spin and is an eigenstate of the AdS mass operator we should ensure that $\Lambda^\sigma$ is symmetric and traceless in the $a_1,\cdots,a_\sigma$ indices. Note that $\Lambda^{2s+1}$ vanishes. See Appendix \ref{AdSEfuncswithKai} for further discussion of $\Lambda^\sigma$ and Appendix \ref{numres} for the explicit construction of a number of eigenstates.

\section{Diagonalizing the AdS mass operator}\label{BC}

In this section we follow the logic given by Metsaev in \cite{Metsaev:1999ui} to construct the eigenfunctions of the AdS mass operator. These eigenfunctions belong to a definite representation of SO(2)$\times$SO($d-2$) and their eigenvalue is determined by the SO($d-2$) spin, an internal spin $p$ to be introduced below, $k$ and $d$.

The irreducible representations of the conformal group SO($2,d$) we consider are constructed using a lowest dimension primary state and its descendants. The primary state is in some irreducible representation of SO($d$) and has a definite dimension $\Delta$. We can decompose the carrier space of the SO($d$) representation into subspaces that are in definite SO(2)$\times$SO($d-2$) representations. A simple example of a state belonging to the $s'$ subspace of an SO($d$) representation with spin $s$, takes the form
\bea
\eta_{s,s'}(\xi^a_q,\rho_q)&=&\left(\prod_{i=1}^k\rho_i^{s_i}\right)\, f_{s'}(\xi^a_q)
\eea
where $f_{s'}(\xi^a_i)$ is a symmetric and traceless (in the $a$ indices) polynomial of degree $s'$ in the $\xi^a_i$ variables and
\bea
\sum_{i=1}^k s_i&=& 2s
\eea
As a simple example we have used
\bea
f_{s'}(\xi^a_i)&=&(\xi^1_i+i\xi^2_i)^{s'}
\eea
when explicitly computing the eigenvalues. Up to an additive constant (see (\ref{simplsmassop})) the AdS mass operator is given by
\bea
O=-{1\over 2}\sum_{i,j=1}^k\sum_{a,b=1}^{d-2}M^{ab}_{ij}M^{ab}_{ij}
\eea
It is straight forward to verify that
\bea
O\,\eta_{s,s'}(\xi^a_q,\rho_q)&=&s'(s'+k(d-2)-d)\,\eta_{s,s'}(\xi^a_q,\rho_q)
\eea
which leads to the following eigenvalues for the AdS mass operator
\bea
A\,\eta_{s,s'}(\xi^a_q,\rho_q)&=&\Big(\kappa_{ss'}^2-{1\over 4}\Big)\,\eta_{s,s'}(\xi^a_q,\rho_q)\label{adsmassop}
\eea
with
\bea
\kappa_{ss'}&=&{(d-2) (k-1)\over 2}+s'-1\label{forkappa}
\eea
The eigenfunctions of the AdS mass operator provide a basis in terms of which we can expand the collective field\footnote{This equation is schematic. For a definite formula we would need to spell out the range of the sums and take care of any multiplicity labels.}
\bea
\Phi(X^+,P^+,X^a,Z,\{\alpha_i\})
&=&\mu(x^a_i,p_i^+)\eta_k(x^+,p_1^+,x^a_1,\cdots,p_k^+,x^a_k)\cr\cr
&=&\sum_{s,s'}\psi_{s,s'}(X^+,P^+,X^a,Z)\eta_{s,s'}(\xi^a_q,\rho_q)
\eea
The eigenvalues of the AdS mass operator fixes the behaviour of the eigenfunction as $Z\to 0$. This follows from the equations of motion for the bulk fields which are given by \cite{Metsaev:1999ui,Metsaev:2003cu,Metsaev:2005ws,Metsaev:2013kaa,Metsaev:2015rda,Metsaev:2022ndg}
\bea
\left(2\partial^+\partial^-+\partial_X^2+\partial_Z^2-{A\over Z^2}\right)\Phi(X^+,P^+,X^a,Z,\xi^a_i,\rho_i) &=&0\label{blkeqnmot}
\eea
In particular, the bulk equation of motion for the $s,s'$ mode is given by
\bea
\left(2{\partial\over\partial X^-}{\partial\over\partial X^+}+\sum_{b=1}^{d-2}{\partial\over \partial X^b}{\partial\over \partial X^b}+{\partial^2\over \partial Z^2}-{\left(\kappa_{ss'}^2-{1\over 4}\right)\over Z^2}\right)\psi_{s,s'}(X^+,X^-,X^a,Z)&=&0
\eea
The normalizable solution to this bulk wave equation is given by \cite{Metsaev:2008fs}
\bea
\psi_{s,s'}(X^+,X^-,X^a,Z)&=&\sqrt{Z}{J_{\kappa_{ss'}}(\hat{q}Z)\over \hat{q}^{\kappa_{ss'}}}g(X^+,X^-,X^a)
\eea
where
\bea
\hat{q}^2&=&2{\partial\over\partial X^-}{\partial\over\partial X^+}+\sum_{b=1}^{d-2}{\partial\over \partial X^b}{\partial\over \partial X^b}
\eea
For small $Z$ we have
\bea
\psi_{s,s'}(X^+,P^+,X^a,Z)\sim Z^{\kappa_{ss'}+{1\over 2}}
\eea
$\psi_{s,s'}$ is dual to a primary operator with so($d-2$) spin $s'$ and a dimension \cite{Metsaev:2013kaa}
\bea
\Delta&=&{d\over 2}+\kappa_{ss'}\,\,=\,\,\Big({d-2\over 2}\Big)k+s'
\eea
This is an example of a GKPW rule for this theory. The above dimension has an appealing interpretation: the dimension is simply the contribution from $k$ free fields plus the dimension coming from the derivatives that produce the spinning state.

In general, we expect a dimension of the form
\bea
\Delta&=&\Big({d-2\over 2}\Big)k+s'+2p
\eea
which comes from an eigenstate of the AdS mass operator with
\bea
\kappa_{ss'}&=&{(d-2) (k-1)\over 2}+s'+2p-1
\eea
In this case $p$ pairs of indices of the derivatives used to produce the primary operator are contracted. We argue in Appendices \ref{AdSEfuncswithKai} and \ref{numres} that $p$ is related to eigenfunctions that are spinning in their internal $i$ indices. Thus, the eigenfunctions of the AdS mass operator that we have obtained are rich enough to reproduce the complete spectrum of primary operators.

\section{Discussion and Conclusions}\label{conclusions}

In this paper, we have continued our series of works \cite{deMelloKoch:2024ewt,deMelloKoch:2024otg,deMelloKoch:2024nfj} that use collective field theory to construct a holographic description of the free matrix conformal field theory. Specifically, we have focused on the construction of rotations in the collective field theory for the complete set of spatial coordinates of the dual bulk AdS spacetime. There are two subtle features to consider: how rotations involving the emergent holographic coordinate are realized and how rotations involving the spatial coordinates that participate in the construction of $X^{\pm}$ are realized.

Our construction of these rotations is remarkably consistent with Metsaev's construction of these operators directly in the gravity \cite{Metsaev:1999ui}. This consistency is another strong confirmation that collective field theory indeed provides a construction of the gravitational dynamics dual to the conformal field theory. Additionally, by using a description of the modes in terms of an SO(2)$\times$SO($d-2$) subgroup of SO($d$), we have derived the eigenfunctions of the AdS mass operator. This has enabled us to derive the GKPW rule for the emergent higher-dimensional theory.

Our construction of SO($d$) has not made contact with the result in \cite{deMelloKoch:2024nfj} which argued that the collective field is defined on the AdS$_{d+1}\times$S$^{k-1}\times$S$^{(d-2)(k-2)}\times$S$^{d-3}$ spacetime. This result was obtained by asking a slightly different question to the question considered in this paper. The paper \cite{deMelloKoch:2024nfj} studied how the collective field is parametrized. In this paper we have constructed the SO($d$) group of rotations. Note that this SO($d$) group is not an isometry of the spacetime since $Z$ is on a different footing to the remaining spatial coordinates. This SO($d$) group is then not expected to be a symmetry of the dynamics and we have explicitly seen that fields with a different number of $Z$ indices have a different behaviour as we send $Z\to 0$.

It is worth stressing that our results, once again, precisely realize the vision outlined in the work of Das and Jevicki \cite{Das:2003vw}.

There are several intriguing directions in which this work should be extended. Exploring the role of the ``internal index $i$'' which distinguishes different fields used to construct the $k$-local collective field is a fascinating question. As we have described at the end of Appendix \ref{AdSEfuncswithKai}, the group of permutations that acts on $i$ will participate in distinguishing different copies of otherwise identical primary operators packaged by the collective field. Clarifying this would allow a detailed and precise description of the complete Fock space of the theory. The detailed description of the primary operators packaged into the trilocal collective field obtained in \cite{deMelloKoch:2024ewt} provides useful data for this exercise.

We have worked strictly at large $N$. It would be interesting to evaluate $1/N$ corrections which are generated by the interaction vertices generated by the Jacobian. Testing that this Jacobian does indeed generate the correct gravitational interactions would be a further highly non-trivial test of the collective field theory approach.

Our focus has been on free conformal field theory; however, the foundational logic and motivation for collective field theory are not inherently restricted to free field theories. In particular, for interacting matrix models, the collective field theory of the invariant loop variables also has ${1\over N}$ as the loop expansion parameter. The arguments of \cite{deMelloKoch:2023ylr} again demonstrate that there will be a redundancy in the collective description, implying that the theory is holographic. Constructing holographic theories through the collective field theory description of interacting conformal field theories presents an interesting open problem.

Beyond interacting conformal field theories, the logic of collective field theory also applies to ordinary gauge theories that do not possess conformal symmetry. Indeed, the original example in \cite{Das:1990kaa} constructs the collective field theory for a theory that is not conformally invariant. Developing the collective field theory for ordinary non-Abelian Yang-Mills theory is a pursuit worth undertaking. Although this is a challenging problem with a long history, one might hope that recent insights gained from AdS/CFT could be instrumental in its resolution.

\begin{center} 
{\bf Acknowledgements}
\end{center}
RdMK is supported by a start up research fund of Huzhou University, a Zhejiang Province talent award and by a Changjiang Scholar award. PR is also supported by the South African Research Chairs Initiative of the Department of Science and Technology and the National Research Foundation. HJRVZ is supported in part by the “Quantum Technologies for Sustainable Development” grant from the National Institute for Theoretical and Computational Sciences of South Africa (NITHECS).

\begin{appendix}

%

\section{GKPW rule for massless fields}\label{GKPW}

The goal of this section is to compare the boundary behaviour of massless fields, that has been worked out in the literature \cite{Metsaev:1999ui,Metsaev:2003cu,Metsaev:2005ws,Metsaev:2013kaa,Metsaev:2015rda,Metsaev:2022ndg} with the boundary behaviour predicted by Section \ref{BC}. This amounts to a study of the spectrum of the AdS mass operator. In general the space $\Lambda$ defined by (\ref{kefuncs}) is invariant under the action of the spin generators $M^{ab}$ and $M^{aZ}$. For massless fields however, $\Lambda$ contains an invariant subspace. In this Appendix we discuss massless fields separately and confirm that their boundary behaviour is correctly reproduced by the collective field theory. Following \cite{Metsaev:1999ui,Metsaev:2003cu,Metsaev:2005ws,Metsaev:2013kaa,Metsaev:2015rda,Metsaev:2022ndg} our approach is based on the bulk equations of motion in light cone gauge. Confirming that the massless fields are eigenfunctions of the AdS mass operator with the correct eigenfunctions establishes that the modes have the correct boundary behaviour. An alternative approach, adopted in \cite{Mintun:2014gua}, is to study the bulk equations of motion in de Donder gauge and then to perform the gauge transformation required to reach light cone gauge.

From Table II of \cite{Metsaev:2013kaa} the AdS mass operator for massless fields in AdS$_{d+1}$ in the\footnote{Recall that in this description the AdS mass operator is diagonal.} so$(2)\oplus$ so$(d-2)$ language is given by
\bea
A&=&\nu^2-{1\over 4}\qquad\qquad \nu\,\,=\,\,s+{d-4\over 2}-N_z\,\,=\,\,s'+{d-4\over 2}
\eea
where we have written the answer in terms of the so$(d-2)$ spin which is given by $s'$. The massless fields are related to the bilocal ($k=2$) collective field. Setting $k=2$ in (\ref{forkappa}) we have
\bea
\kappa_{ss'}&=&s'+{d-4\over 2}
\eea
Since by (\ref{adsmassop}) we have $A=\kappa_{ss'}^2-{1\over 4}$ we see that our eigenfunctions correctly describe the massless fields in AdS$_{d+1}$.

\section{AdS Mass Operator eigenfunctions}\label{AdSEfuncswithKai}

In this Appendix we explore the description of the eigenfunctions of the AdS mass operator given in equation (\ref{kefuncs}). By an SO($d$) representation of spin $s$ we mean the irreducible representation of SO($d$) spanned by traceless and symmetric tensors with $s$ indices and each index takes on $d$ values. There is an SO(2)$_{\rm l.c.}\times$SO($d-2$) subgroup of SO($d$). Here SO($d-2$) rotations mix the coordinates $X^a$ transverse to the light cone and SO(2)$_{\rm l.c.}$ mixes $Z$ and $X^{d-1}$, where $X^{d-1}$ is the coordinate that is combined with time $T$ to define $X^\pm$. The description provided by (\ref{kefuncs}) puts the states into definite irreducible representations of SO($d-2$), labelled by a definite eigenvalue of the single generator in SO(2)$_{\rm l.c.}$, which is given by ${\sf d}$. Introduce the state $|{\bf 0}\rangle$, defined by
\bea
|{\bf 0}\rangle &=& \prod_{i=1}^k\rho_i^{l_i}
\eea
where
\bea
2s&=&\sum_{i=1}^k l_i
\eea
for the description of an SO($d$) representation of spin $s$. We construct the remaining states of the representation by applying the ${\sf k}^a_i$ operators to $|{\bf 0}\rangle$ as follows
\bea
|a_1,\cdots,a_n\rangle&\equiv&{\sf k}_i^{a_1}{\sf k}_i^{a_2}\cdots{\sf k}_i^{a_n} |{\bf 0}\rangle
\label{defstaten}
\eea
Since the ${\sf k}^a_i$ operators commute, the state obtained is automatically symmetric in it's $a_j$ indices. In addition, since it is not in general traceless, this state is in general a reducible representation of the so($d-2$). It is simple to verify that these states are eigenfunctions with a definite so(2)$_{\rm l.c.}$ spin
\bea
d|a_1,\cdots,a_n\rangle&=&(n-s)\,|a_1,\cdots,a_n\rangle
\eea
For odd $n$ the space spanned by the states $\{|a_1,\cdots,a_n\rangle\}$ can be decomposed into SO($d-2$) irreducible representations as follows
\bea
{\rm Span}\{|a_1,\cdots,a_n\rangle\}&=& {\bf n}\oplus {\bf n-2}\oplus {\bf n-4}\oplus\cdots\oplus {\bf 1}
\eea
while for even $n$ the decomposition is
\bea
{\rm Span}\{|a_1,\cdots,a_n\rangle\}&=& {\bf n}\oplus {\bf n-2}\oplus {\bf n-4}\oplus\cdots\oplus {\bf 0}
\eea
where ${\bf r}$ denotes the irreducible representation of SO($d-2$) with spin $r$. The states belonging to ${\bf r}$ obey
\bea
-{1\over 2}\sum_{a,b=1}^{d-2}\,J^{ab}J^{ab}|a_1,\cdots,a_n;{\bf r}\rangle&=&r(r+d-4)|a_1,\cdots,a_n;{\bf r}\rangle
\eea
and they are eigenstates of the AdS mass operator
\bea
A|a_1,\cdots,a_n;{\bf r}\rangle&=&\left(\left({(d-2) (k-1)\over 2}+r-1\right)^2-{1\over 4}\right)|a_1,\cdots,a_n;{\bf r}\rangle
\eea

\begin{table}[h]
\centering
\begin{tabular}{|c|c|c|c|c|}
\hline
\,\,\,\, $n$\,\,\,\, & \,\,\,\,so(2)$_{l.c.}=\lambda_{{\rm so(2)}_{l.c.}}$\,\,\,\, & \,\,\,\,so($d-2$)$_{\perp}$\,\,\,\, &\,\,\,\,{\rm dim}($d=4$)\,\,\,\,&\,\,\,\,{\rm dim}($d=5$)\,\,\,\,\\
\hline
0 & $-4$ & ${\bf 0}$ & 1 &1\\
\hline
1 & $-3$ & ${\bf 1}$ & 2 &3\\
\hline
2 & $-2$ & ${\bf 0}\oplus{\bf 2}$ & 3 &6\\
\hline
3 & $-1$ & ${\bf 1}\oplus{\bf 3}$ & 4 &10\\
\hline
4 & $0$  & ${\bf 0}\oplus{\bf 2}\oplus{\bf 4}$ & 5 &15\\
\hline
5 & $1$  & ${\bf 1}\oplus{\bf 3}$ & 4 &10\\
\hline
6 & $2$  & ${\bf 0}\oplus{\bf 2}$ & 3 &6\\
\hline
7 & $3$  & ${\bf 1}$ & 2 &3\\
\hline
8 & $4$  & ${\bf 0}$  & 1 &1\\
\hline
\end{tabular}
\caption{This table shows the so(2)$\oplus$so($d-2$) description of the spin 4 representation of so($d$). The first column records the integer $n$ appearing in (\ref{defstaten}). The second column records the eigenvalue of the single generator of so(2)$_{\rm l.c.}$. The third column records the irreducible representations of so($d-2$) that appear at the given $n$. The sum of the dimensions of these irreducible representations at a given $n$ are recorded in column four (for $d=4$) and in column five (for $d=5$). The sum of entries in the fourth column gives 25 which is the correct answer for the dimension of the spin 4 representation of SO(4). The sum of entries in the fifth column gives 55 which is the correct answer for the dimension of the spin 4 representation of SO(5).}\label{Sis4}
\end{table}

The eigenvalues of the AdS mass operator fixes the behaviour of the eigenfunction as $Z\to 0$. As explained in Section \ref{BC} the normalizable solution the bulk field behaves as $Z^{{1\over 2}+\kappa}$ where we write the eigenvalue of the AdS mass operator as $\kappa^2-{1\over 4}$. We have described a concrete example of the spin 4 representation of SO($d$) in Table \ref{Sis4}.

The fact that we know the so(2)$\oplus$so($d-2$) transformation properties of the eigenfunctions implies that we know the values of tensor indices of the corresponding bulk field. In a light cone gauge fixing we have eliminated all $+$ indices (by our gauge choice) and we have eliminated all $-$ indices by solving the constraints. Consequently we are interested in bulk fields whose indices take values in  $\{Z,X^a\}$. The bulk field with $w$ indices equal to $Z$ and the remaining indices chosen from the $X^a$'s is constructed from states with $|\lambda_{{\rm so(2)}_{l.c.}}|=w$.

The case that $d=3$, corresponding to AdS$_4$, deserves special attention. In this case there is only one direction transverse to the light cone (denoted simply as $X$) and the SO($d-2$) group is replaced by $\mathbb{Z}_2$. The two possible representations are $\pm$ depending on whether the tensor is even or odd under parity $X\to -X$. These are both 1 dimensional. The counting works out for this case as it did for $d>3$. For example, in Table \ref{Sis2} we have  considered the spin 2 representation of SO(3).

\begin{table}[h]
\centering
\begin{tabular}{|c|c|c|}
\hline
\,\,\,\, $n$\,\,\,\, & \,\,\,\,so(2)$_{l.c.}=\lambda_{{\rm so(2)}_{l.c.}}$\,\,\,\, & \,\,\,\,\,\,\,\,$\mathbb{Z}_2$\,\,\,\,\,\,\,\,\\
\hline
0 & $-2$ & $+$\\
\hline
1 & $-1$ & $-$\\
\hline
2 & $0$ & $+$\\
\hline
3 & $1$ & $-$\\
\hline
4 & $2$  & $+$\\
\hline
\end{tabular}
\caption{This table shows the so(2)$\oplus\mathbb{Z}_2$ description of the spin 2 representation of so(3). The first column records the integer $n$ appearing in (\ref{defstaten}). The second column records the eigenvalue of the single generator of so(2)$_{\rm l.c.}$. The third column records the parity of the $\mathbb{Z}_2$ representation. There are a total of 5 states which equals the dimension of the spin 2 representation of so(3).}\label{Sis2}
\end{table}

A notable gap in our analysis is related to the index $i$ that distinguishes the different fields appearing in the $k$-local collective field. In the construction (\ref{defstaten}) that we have explored in this Appendix we have restricted $i$ to a single copy. This is certainly not the only choice. To motivate the discussion that follows, recall that the most general primary operator constructed from $k$ fields that we can consider takes the form
\bea
\sum_{n_1 n_2 \cdots n_k;\beta}c_{n_1 n_2 \cdots n_k;\beta}\Tr\Big((\epsilon\cdot\partial)^{n_1}\partial_{f_1}\phi(x)\,(\epsilon\cdot\partial)^{n_2}\partial^{f_1}\phi(x)\,\cdots (\epsilon\cdot\partial)^{n_k}\partial^{f_p}\phi(x)\Big)
\eea
where $\epsilon^\mu$ is a polarization tensor used to simplify the above formula, $\beta$ labels terms for which the derivatives with contracted indices act on different fields, $n_1+n_2+\cdots+n_k=s'$ and $c_{n_1 n_2 \cdots n_k;\beta}$ are some numbers chosen so that the above operator is primary. There are $p$ pairs of derivatives which are contracted. The operator is symmetric and traceless in the remaining $s'$ free indices which have been contracted with $\epsilon^\mu$ above , i.e. it has so($d-2$) spin $s'$. The dimension of this operator, since we are studying the free field theory, is
\bea
\Delta&=&\Big({d-2\over 2}\Big)k+s'+2p
\eea
From the results in \cite{Metsaev:2013kaa}, to obtain an operator with this dimension we need an eigenfunction of the AdS mass operator that has
\bea
\kappa_{ss'}&=&{(d-2) (k-1)\over 2}+s'+2p-1\label{withp}
\eea
The AdS operator is sensitive to both the spin in the space indices $a$ and in the internal indices $i$. The spacetime angular momentum on the other hand is sensitive only to the spin in the space indices. Thus, to obtain a non-zero value for $p$ we can consider an eigenfunction that is spinning in the internal indices. We give a concrete example of this at the end of Appendix \ref{numres}.

Finally, note that operators belonging to $\mathbb{Z}_k$ which act as a permutation\footnote{Since we are considering a permutation, one might also study the group $S_k$. However, this is not in general a symmetry of the $k$-local collective field. Thanks to cyclicity of the trace $\mathbb{Z}_k$ is indeed a symmetry.} on this index commute with the AdS mass operator and spacetime rotations, so they give another state with exactly the same quantum numbers. Thus, this $\mathbb{Z}_k$ group belongs to the commutant of SO(2,$d$) and it will play some role in oragnizing the representations of SO(2,$d$) (i.e. the primary operators) packaged in the $k$-local collective field. It would be interesting to work this out in detail.

\section{Numerical Results}\label{numres}

In this Appendix we explicitly describe some of the eigenfunctions of the AdS mass operator for a few carefully chosen examples. Our main motivation for doing this is to point out some subtleties associated with the treatment of the index $i$ that distinguishes different fields used to produce the $k$-local collective field.

\subsection{so($d-2$) spin 1 with $k=3$ and $d-2=2$}

The spin 1 representation has a total of 4 states. They are given by
\bea
\psi_0&=&\rho_1\cr\cr
 \psi_1&=&{\sf k}^1_1\rho_1\qquad\qquad
\psi_2\,\,=\,\,{\sf k}^2_1\rho_1\cr\cr
\psi_3&=&{\sf k}^1_1{\sf k}^1_1\rho_1\,\,=\,\,{\sf k}^2_1{\sf k}^2_1\rho_1
\eea
Note that ${\sf k}^1_1{\sf k}^2_1\rho_1=0$. It is simple to verify that
\bea
-{1\over 2}\sum_{a,b=1}^2 J^{ab}J^{ab}\psi_\alpha&=&0\qquad\qquad \alpha=0,3\cr\cr
-{1\over 2}\sum_{a,b=1}^2 J^{ab}J^{ab}\psi_\beta&=&\psi_\beta\qquad\qquad\beta=1,2
\eea
We also have
\bea
A\psi_\alpha&=&\Big(1^2-{1\over 4}\Big)\psi_\alpha\qquad\qquad \alpha=0,3\cr\cr
A\psi_\beta&=&\Big(2^2-{1\over 4}\Big)\psi_\beta\qquad\qquad\beta=1,2
\eea

\subsection{so($d-2$) spin 2 with $k=3$ and $d-2=2$}

The spin 2 representation has a total of 9 states. They are given by
\bea
\psi_0&=&\rho_1^2\cr\cr 
\psi_1&=&{\sf k}^1_1\rho_1^2\qquad\qquad
\qquad\psi_2\,\,=\,\,{\sf k}^2_1\rho_1^2\cr\cr
\psi_3&=&{\sf k}^1_1{\sf k}^2_1\rho_1^2\qquad\qquad
\psi_4\,\,=\,\,({\sf k}^1_1{\sf k}^1_1-{\sf k}^2_1{\sf k}^2_1)\rho_1^2\cr\cr
\psi_5&=&\sum_{a=1}^2\sum_{i=1}^3{\sf k}^a_i{\sf k}^a_i\rho_1^2\cr\cr
\psi_6&=&{\sf k}^1_1{\sf k}^2_1{\sf k}^2_1\rho_1^2\qquad\qquad
\psi_7\,\,=\,\,{\sf k}^2_1{\sf k}^2_1{\sf k}^2_1\rho_1^2\cr\cr
\psi_8&=&{\sf k}^1_1{\sf k}^1_1{\sf k}^1_1{\sf k}^1_1\rho_1^2
\eea
It is simple to verify that
\bea
-{1\over 2}\sum_{a,b=1}^2 J^{ab}J^{ab}\psi_\alpha&=&0\qquad\qquad \alpha=0,5,8\cr\cr
-{1\over 2}\sum_{a,b=1}^2 J^{ab}J^{ab}\psi_\beta&=&\psi_\beta\qquad\qquad\beta=1,2,6,7\cr\cr
-{1\over 2}\sum_{a,b=1}^2 J^{ab}J^{ab}\psi_\gamma&=&4\psi_\gamma \qquad\qquad\gamma=3,4
\eea
We also have
\bea
A\psi_\alpha&=&\Big(1^2-{1\over 4}\Big)\psi_\alpha\qquad\qquad \alpha=0,5,8\cr\cr
A\psi_\beta&=&\Big(2^2-{1\over 4}\Big)\psi_\beta\qquad\qquad\beta=1,2,6,7\cr\cr
A\psi_\gamma&=&\Big(3^2-{1\over 4}\Big)\psi_\gamma\qquad\qquad\gamma=3,4
\eea
The states $\psi_0$, $\psi_5$ and $\psi_8$ are all singlets of so($d-2$) and they all have the same eigenvalue for the AdS mass operator. The sum over $i$ in the construction of $\psi_5$ is not needed to produce a singlet of so($d-2$), but it is needed to obtain an eigenstate of the AdS mass operator. This is one interesting lesson from these numerical results. The pairs of states $\psi_1,\psi_2$ and $\psi_6,\psi_7$ are both spin 1 doublets of so($d-2$). Finally, $\psi_3,\psi_4$ is a spin 2 doublet of so($d-2$).

\subsection{Spinning on the internal indices with $k=3$ and $d-2=2$}

In this subsection we consider a very simple example of a state that is spinning in the internal indices. The state
\bea
\psi&=&\sum_{a=1}^2{\sf k}^a_1{\sf k}^a_2\rho_1^2
\eea
obeys
\bea
-{1\over 2}\sum_{a,b=1}^2 J^{ab}J^{ab}\psi&=&0\qquad\qquad
A\psi\,\,=\,\,\Big(3^2-{1\over 4}\Big)\psi
\eea
It is a solution with eigenvalue given by (\ref{withp}) with $s'=0$ and $p=1$.

\end{appendix}

\end{document}